\let\pdfoutput\defined

\documentclass{PoS}
\usepackage{amsmath, amsfonts, bbm, dsfont, mathrsfs,amssymb}

\newlength{\closercaption}
\title{%
\vspace*{-1cm}
\begin{minipage}{\textwidth}
\begin{flushright}
\texttt{\footnotesize
PoS(LAT2007)036\\%
DESY-07-157\\%
BNL-HET-07/15\\%
}
\end{flushright}
\end{minipage}\\[15pt]
Stout Smearing for Twisted Mass Fermions}

\ShortTitle{Stout Smearing for Twisted Mass Fermions}
\author{\speaker{Karl Jansen}\\
NIC/DESY Zeuthen, Platanenallee 6, D-15738 Zeuthen, Germany\\
E-mail: \email{karl.jansen@desy.de}}

\author{Craig McNeile\\
Dept. of Physics and Astronomy, Kelvin Build., University of Glasgow, Glasgow G12 8QQ, UK\\
E-mail: \email{c.mcneile@physics.gla.ac.uk}}

\author{Istv\'an Montvay\\
Deutsches Elektronen-Synchrotron DESY, Notkestr.~85, D-22607 Hamburg, Germany\\
E-mail: \email{istvan.montvay@desy.de}}

\author{Chris Richards\\
Theoretical Physics Div., Dept.\ of Mathematical Sciences, University of Liverpool,\\ Liverpool L69 7ZL, UK\\
E-mail: \email{c.m.richards@liverpool.ac.uk}}

\author{Enno E. Scholz\\
Physics Department, Brookhaven National Laboratory, Upton, NY, 11973, USA\\
E-mail: \email{scholzee@quark.phy.bnl.gov}}

\author{Carsten Urbach\thanks{Current address: Humboldt-Universit\"at zu Berlin, Institut f\"ur Physik, Newtonstr.\ 15, 12489 Berlin, Germany}\\
Theoretical Physics Div., Dept.\ of Mathematical Sciences, University of Liverpool,\\ Liverpool L69 7ZL, UK\\
E-mail: \email{Carsten.Urbach@liverpool.ac.uk}}

\author{Urs Wenger\\
Institute for Theoretical Physics, ETH Z\"urich, CH-8093 Z\"urich, Switzerland\\
E-mail: \email{wenger@phys.ethz.ch}}
\abstract{%
 The effect of Stout smearing is investigated in numerical simulations with
 twisted mass Wilson quarks.
 The phase transition near zero quark mass is studied on
 $12^3\times24$, $16^3\times32$ and $24^3\times48$ lattices at lattice spacings
 $a \simeq 0.1$ -- $0.125\, {\rm fm}$.}
\FullConference{The XXV International Symposium on Lattice Field Theory\\
		 July 30 - August 4 2007\\
		 Regensburg, Germany}
\begin{document}

\hyphenation{pla-quet-te }

\newcommand{\rmi}{{\rm i}}

\newcommand{\be}{\begin{equation}}                                              
\newcommand{\ee}{\end{equation}}                                                
\newcommand{\half}{\frac{1}{2}}                                                 
\newcommand{\rar}{\rightarrow}                                                  
\newcommand{\lar}{\leftarrow}
\newcommand{\LCB}{\raisebox{-0.3ex}{\mbox{\LARGE$\left\{\right.$}}}
\newcommand{\RCB}{\raisebox{-0.3ex}{\mbox{\LARGE$\left.\right\}$}}}
\newcommand{\U}{\sf U}
\newcommand{\SU}{\sf SU}

\newcommand{\partialL}{\stackrel{\leftarrow}{\partial}}
\newcommand{\partialR}{\stackrel{\rightarrow}{\partial}}

\newcommand{\bteq}[1]{\boldmath$#1$\unboldmath}
\newcommand{\eins}{\mathds 1}


\def\nicefrac#1#2{\leavevmode\kern.1em\raise.5ex\hbox{\the\scriptfont0 #1}\kern-.1em/\kern-.15em\lower.25ex\hbox{\the\scriptfont0 #2}}

\newcommand{\R}{\boldsymbol{\mathbb{R}}}

\newcommand{\C}{\boldsymbol{\mathbb{C}}}

\newcommand{\N}{\boldsymbol{\mathbb{N}}}

\newcommand{\Z}{\boldsymbol{\mathbb{Z}}}


\makeatletter
  \newcommand{\fslash}[2][0mu]{%
    \mathchoice
      {\fsl@sh\displaystyle{#1}{#2}}%
      {\fsl@sh\textstyle{#1}{#2}}%
      {\fsl@sh\scriptstyle{#1}{#2}}%
      {\fsl@sh\scriptscriptstyle{#1}{#2}}}
  \newcommand{\fsl@sh}[3]{%
    \m@th\ooalign{$\hfil#1\mkern#2/\hfil$\crcr$#1#3$}}
\makeatother





\newcommand{\tr}{\mathrm{tr}}

\newcommand{\Tr}{\mathrm{Tr}}

\newcommand{\str}{\mathrm{str}}

\newcommand{\sTr}{\mathrm{sTr}}

\newcommand{\Real}{\mathrm{Re}}

\newcommand{\Imag}{\mathrm{Im}}

\newcommand{\sci}[2]{#1\cdot10^{#2}}

\begin{figure}
\begin{center}
\hfill
\includegraphics[angle=90, width=0.53\textwidth]{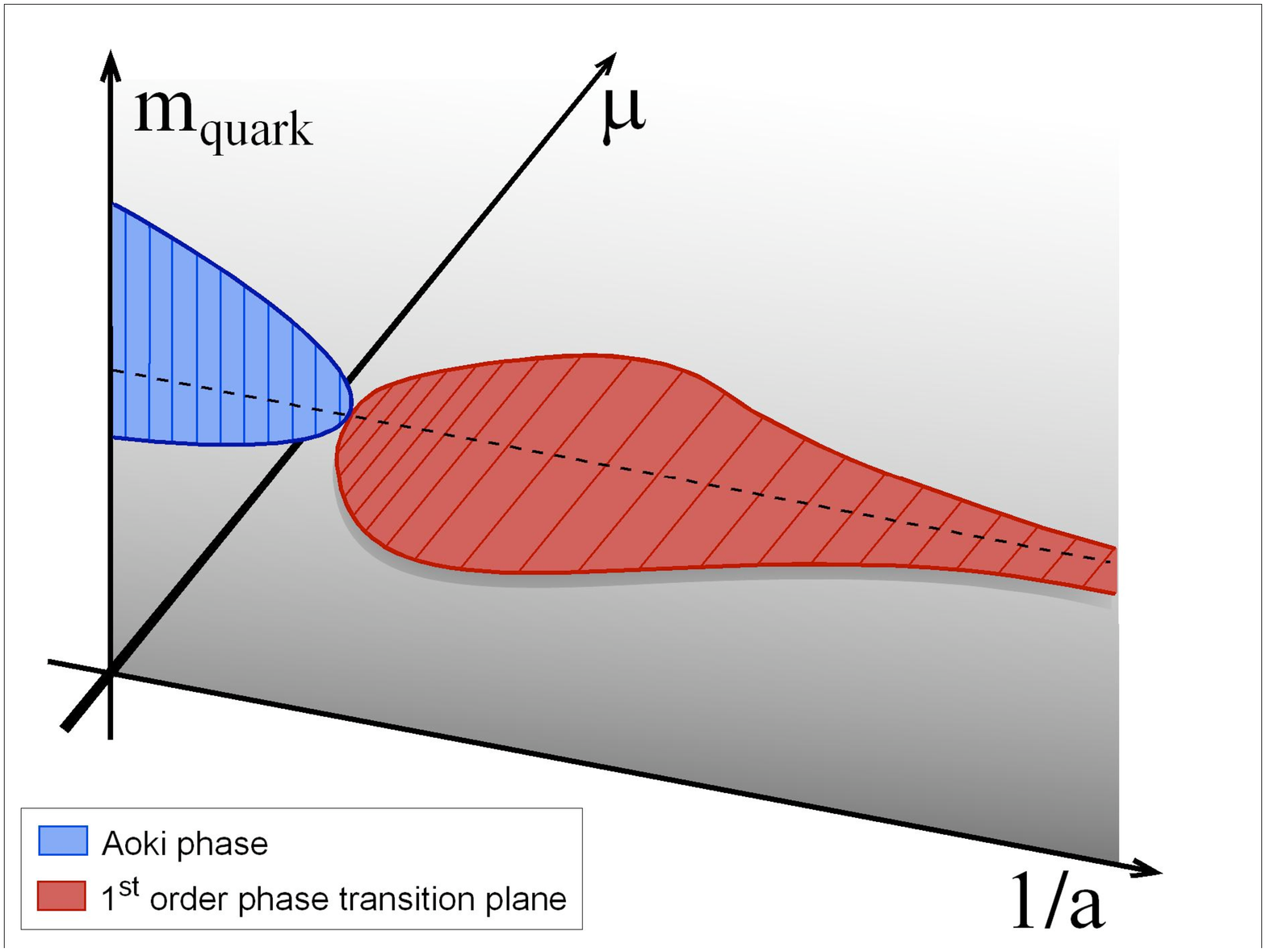}
\hfill
\includegraphics[width=0.31\textwidth]{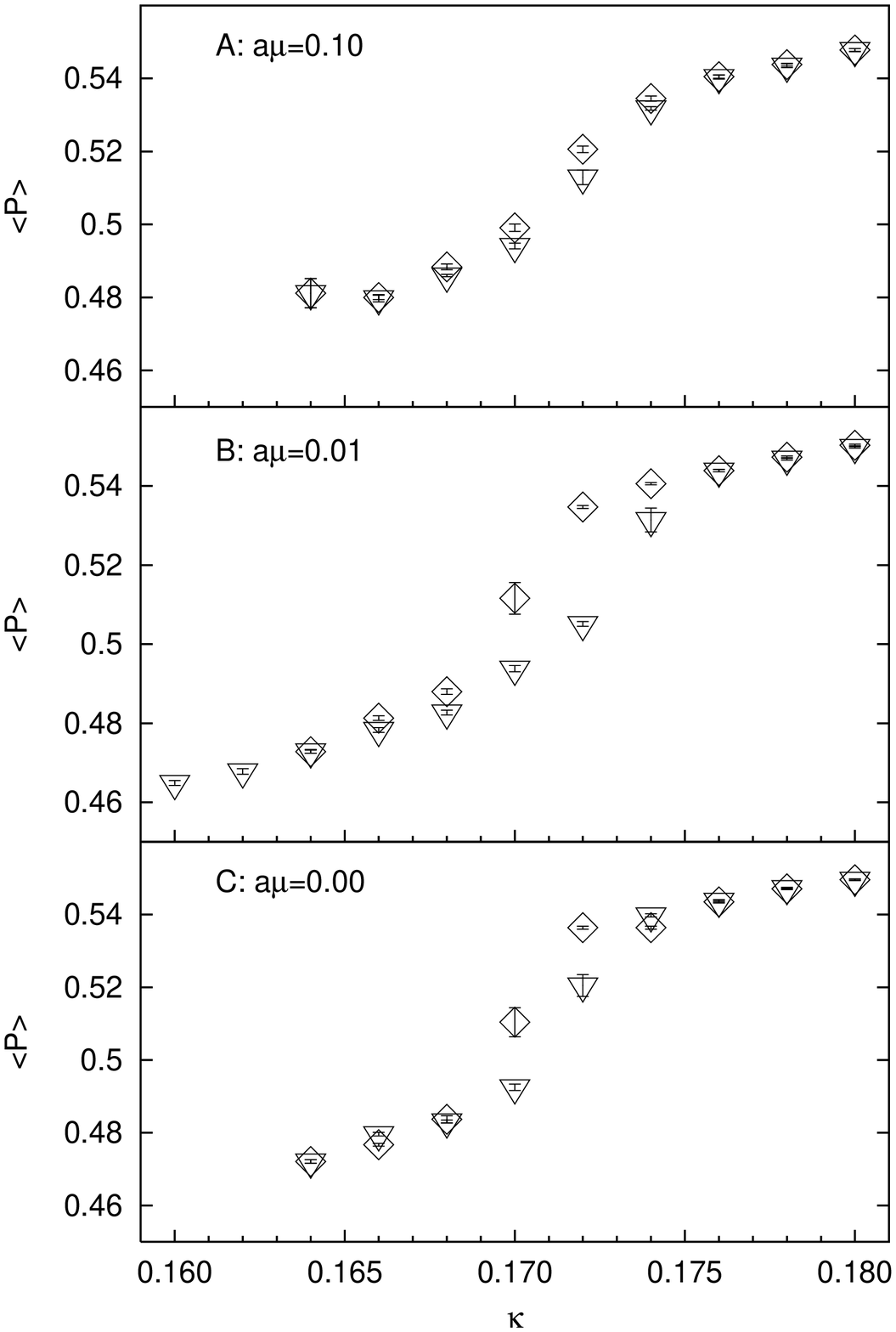} 
\hfill$\phantom{.}$
\end{center}
\vspace*{\closercaption}
\caption{\label{fig:phasediag}\textit{Left panel:} The phase diagram with Wilson fermions. \textit{Right panel:} Thermal cycles with $a\mu=0.1$, $0.01$, and without a twisted mass term on $8^3\times16$ lattices at $\beta=5.2$ (Wilson gauge action), see \cite{Farchioni:2004us}.}
\end{figure}

 The phase structure of Wilson fermions with twisted mass ($\mu$) has been
 investigated in \cite{Farchioni:2004us, Farchioni:2005tu}.
 As it is explained there, the observed first order phase transition limits
 the minimal pion mass which can be reached in simulations at a given lattice
 spacing: $m_\pi^{\rm min}\simeq\mathcal{O}(a)$.
 The phase structure is schematically depicted in the left panel of
 Fig.~\ref{fig:phasediag}.
 The phase transition can be observed in simulations with twisted mass fermions,
 for instance, as a ``jump'' or even metastabilities in the average plaquette
 value as a function of the hopping parameter ($\kappa$).
 (The right panel of Fig.\ \ref{fig:phasediag} shows thermal cycles to
 demonstrate this phenomenon.)

 One possibility to weaken the phase transition and therefore allow for lighter
 pion masses at a given lattice spacing is to use an improved gauge action
 like the DBW2, Iwasaki, or tree-level Symanzik (tlSym) improved gauge action instead
 of the simple Wilson gauge action.
 This has been successfully demonstrated in
 \cite{Farchioni:2004fs,Farchioni:2005bh,Boucaud:2007uk}.

 Here we report on our attempts to use a smeared gauge field in the fermion
 lattice Dirac operator to further reduce the strength of the phase transition.
 This is relevant in simulations with
 $N_f=2+1+1$ $(u,d,s,c)$ quark flavours \cite{Chiarappa:2006ae} where the
 first order phase transition becomes stronger compared to $N_f=2$ simulations.
 The main impact of the above mentioned improved gauge actions on the gauge
 fields occuring in simulations is to suppress short range fluctuations
 (``dislocations'') and the associtated ``exceptionally small'' eigenvalues
 of the fermion matrix.
 The same effect is expected from smearing the gauge field links in the
 fermion action.
 The cumulated effect of the improved gauge action and smeared links should
 allow for a smaller pion mass at a given lattice spacing and volume.
 Our choice is the Stout smearing procedure as introduced in
 \cite{Morningstar:2003gk}, since it can easily be implemented in the Hybrid
 Monte Carlo (HMC) based updating algorithms we are currently using. 

 One should keep in mind that a possible caveat of this procedure is
 ``oversmearing'', i.e., removing too many small eigenvalues by applying too
 many smearing steps and/or using a too high value for the smearing parameter---because not every small eigenvalue is ``unphysical''.
 In addition, after many smearing steps the fermion action can become too
 delocalised which can lead to an unwanted slowing down of the approach to
 the continuum limit.
 In order to avoid this caveat we choose to work with only one step of
 very mild Stout smearing. Moreover we keep these smearing parameters
 fixed as we change the lattice spacing.

 In Section \ref{sec:stout} we will shortly review the smearing procedure and
 the twisted mass formulation, as well as some details concerning the used updating algorithms.
 Section\ \ref{sec:numSim} is devoted to the presentation of the results of our
 numerical simulations using $N_f=2$ and $N_f=2+1+1$ flavours of twisted mass quarks.
%
\section{\label{sec:stout}Stout smearing and twisted mass fermions}
%
%
\subsection{\label{subsec:smear}Analytic smearing for SU(3) link variables}

 To have a smearing procedure which is analytic in the unsmeared link variables
 is an essential feature, if one is to use the smeared variables in an updating
 scheme like HMC based algorithms requiring the calculation of the derivative
 (or force) with respect to the unsmeared link variables.
 The Stout smearing procedure as introduced by Morningstar and Peardon in
 \cite{Morningstar:2003gk} was designed to meet this requirement.
 We will briefly describe it in the following but will limit ourselves to the
 case of ${\rm SU}(3)$.
 For more details we refer to the cited work.

 The $(n+1)^{\rm th}$ level of Stout smeared gauge links is obtained iteratively
 from the $n^{\rm th}$ level by
\be U_\mu^{(n+1)}(x)\;=\;e^{\rmi\,Q_\mu^{(n)}(x)}\,U_\mu^{(n)}(x). \ee
 In the following we will refer to the unsmeared (``thin'') gauge field as
 $U_\mu\equiv U_\mu^{(0)}$, while the maximally smeared (``thick'') gauge field
 for $N$-level Stout smearing will be denoted as
 $\widetilde{U}_\mu\equiv U_\mu^{(N)}$.
 The ${\rm SU}(3)$ matrices $Q_\mu$ are defined via the staples $C_\mu$:
\begin{eqnarray}
	Q_\mu^{(n)}(x) &=& \frac{\rmi}2\Big[U^{(n)}_\mu(x){C_\mu^{(n)}}^\dagger(x)
 - {\rm h.c.}\Big]\,-\,\frac{\rmi}{6}\Tr\Big[U^{(n)}_\mu(x){C_\mu^{(n)}}^\dagger(x)
 - {\rm h.c.}\Big]\,,\\
	C_\mu^{(n)} &=& \sum_{\nu\neq\mu}\,\rho_{\mu\nu}\,
\Big(U_\nu^{(n)}(x)U_\mu^{(n)}(x+\hat\nu){U_\nu^{(n)}}^\dagger(x+\hat\mu)
\nonumber\\
		&& \;\;\;
+{U_\nu^{(n)}}^\dagger(x-\hat\nu)U_\mu^{(n)}(x-\hat\nu)U_\nu^{(n)}(x-\hat\nu+\hat\mu)
\Big)\,,
\end{eqnarray}
 where in general $\rho_{\mu\nu}$ is the smearing matrix.
 In our numerical simulations we used exclusively isotropic 4-dimensional smearing,
 i.e., $\rho_{\mu\nu}=\rho$.

The thick gauge field will only be used in the fermion operator, cf.\ Eq.~(\ref{eq:NplusR}). Therefore the usage of smeared links is nothing else but a different discretization of the covariant derivative operator on the lattice. For the gauge part of the action the thin gauge field still is the relevant one.

\subsection{\label{subsec:tmAct}Twisted mass fermion action}

 The notations in this subsection follow Ref.\ \cite{Chiarappa:2006ae}.
 (For details we refer to this work.)
 We performed simulations with one light doublet $(u,d)$ of twisted mass Wilson
 fermions (only using the fermion matrix $Q^{(\chi)}_{l}$).
 Later on we shall add a second doublet for the heavier quarks $(c,s)$, where the
 masses are non-degenerate due to the addition of an extra mass term
 (cf. \cite{Frezzotti:1999vv,Frezzotti:2003xj}). The fermion action then reads
  \begin{eqnarray}
    S^{\rm fermion} &=& \sum_{x,y}\,
\Big(\overline{\chi}_{l,x}\,Q^{(\chi)}_{l,xy}\,\chi_{l,y}\,
+\,\overline{\chi}_{h,x}\,Q^{(\chi)}_{h,xy}\,\chi_{h,y} \Big)\,,
\\
Q^{(\chi)}_{l}  &=& \mu_{\kappa_l}\,+\,\rmi\gamma_5\tau_3a\mu_l\,+\,(N+R)_{xy}\,, 
\\
\label{eq:tm_heavyDoublet}
Q^{(\chi)}_{h} &=& \mu_{\kappa_h}\,+\,\rmi\gamma_5\tau_1a\mu_\sigma\,
+\,\tau_3a\mu_\delta\,+(N+R)_{xy}\,,
\\\label{eq:NplusR}
(N+R)_{xy} &=&
-\frac12\sum_{\mu=\pm1}^{\pm 4}\,\delta_{x,y+\hat\mu}\,\widetilde{U}_{\mu}(y)\,
\Big(\gamma_\mu+r\Big)\,,
\end{eqnarray}
 where $\mu_{\kappa_X} = 1/(2\kappa_X)$ is the untwisted mass and
 $a\mu_l$, $a\mu_h$, and $a\mu_\delta$ are the twisted mass terms in the light
 and heavy doublet and the split mass term, respectively.

\subsection{\label{subsec:alg}Algorithms}

 We used two different algorithms with independent implementations of the Stout smearing
 routines to be able to cross-check our results.
 The first algorithm is the HMC algorithm with multiple time scale integration and mass
 preconditioning as described in \cite{Urbach:2005ji}.
 In that case, the smearing routines were taken from the \textsc{Chroma}-code package
 \cite{Edwards:2004sx} and a chronological inverter was included, too.

 Since this algorithm only allows to simulate an even number of fermion flavours (also
 excluding the case of a split doublet as described in Sec.~\ref{subsec:nf211}), as
 a preparation for the $N_f=2+1+1$ simulations, we also added Stout smearing routines to
 our existing Polynomial HMC (PHMC) \cite{deForcrand:1996ck,Frezzotti:1997ym} update code, where we perform one stochastic
 correction step at the end of a trajectory.
 For details on the implementation of the PHMC, cf.~\cite{Montvay:2005tj,Scholz:2006hd}.
 We used trajectory lengths of $2\times0.35$ to $3\times0.35$ and determinant breakup
 of $n_B=2$.

\section{\label{sec:numSim}Numerical simulations}
\subsection{\label{subsec:nf2}$N_f=2$}

 In all of the simulations presented here, we used the tree-level Symanzik 
 improved gauge action on either $12^3\times24$, $16^3\times32$, or
 $24^3\times48$ lattices.
 We compare results obtained using one level of Stout smearing ($N=1$) with
 $\rho=0.1$ or  $\rho=0.125$ to simulations without smearing of the link
 variables in the Wilson twisted mass fermion action.
 Our choice of mild smearing (smearing only once with a small parameter)
 should guarantee that the fermion action remains well localized on physical
 scales even on relatively coarse lattice spacings.

\begin{figure}
\begin{center}
\includegraphics[angle=-90, width=.43\textwidth]{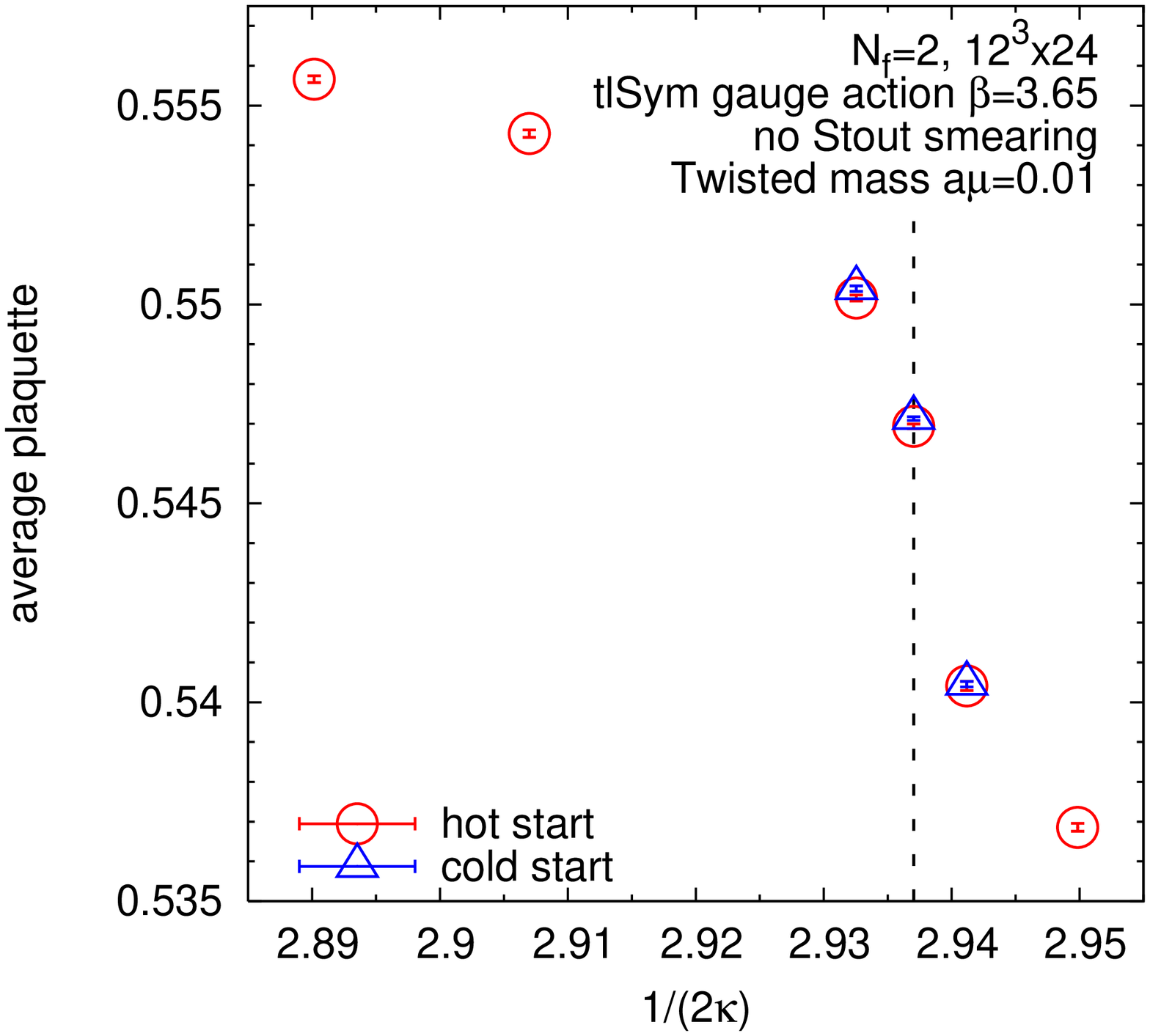}%
\includegraphics[angle=-90, width=.43\textwidth]{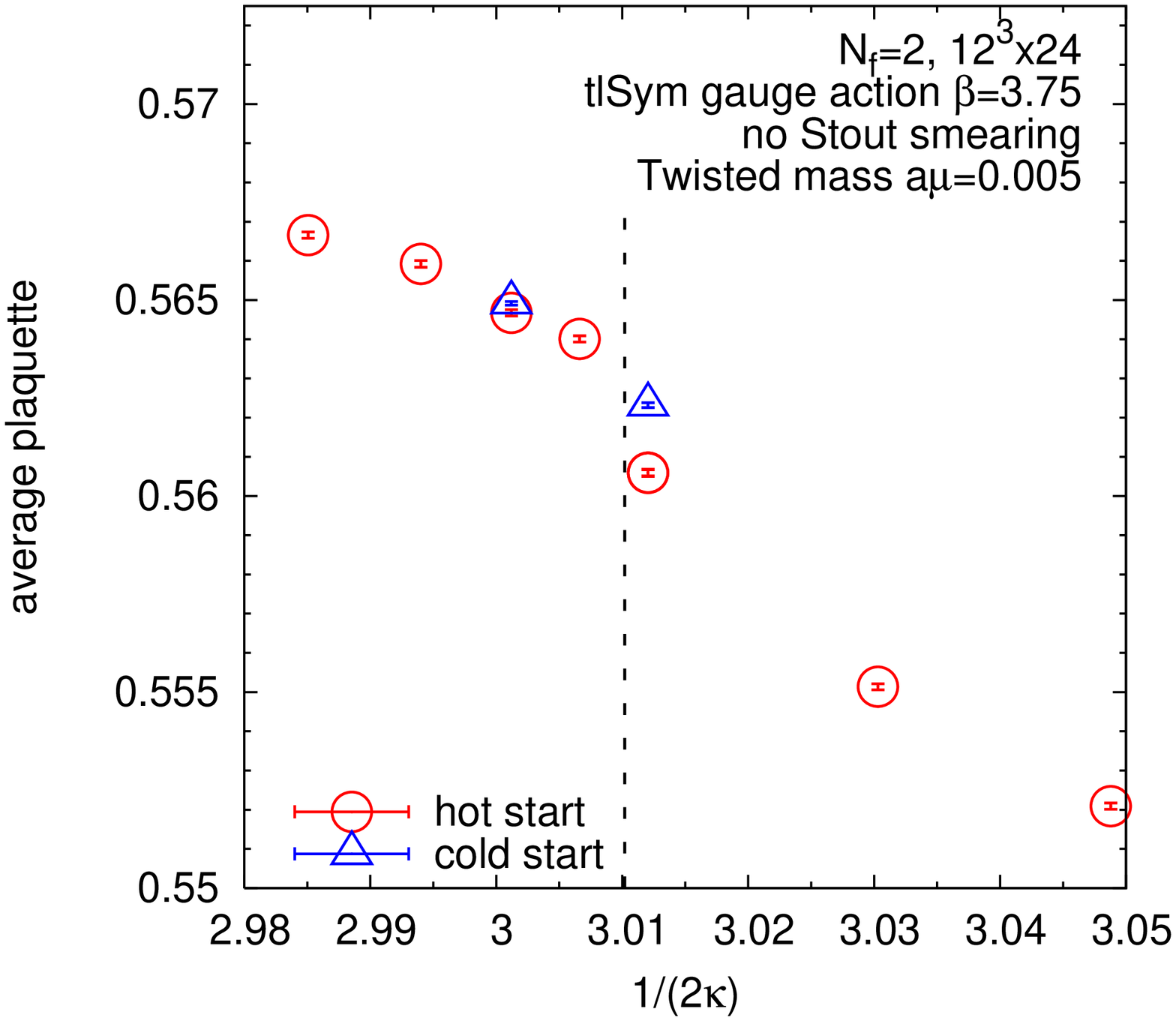}%
\end{center}
\vspace*{\closercaption}
\caption{\label{fig:nf2_plaq_noStout} Average plaquette value without Stout smearing with hot \textit{(red circles)} and cold starts \textit{(blue triangles)} at two different values for the gauge coupling $\beta$ and twisted mass $a\mu$.}  
\end{figure}

\begin{figure}
\begin{center}
\includegraphics[angle=-90, width=.43\textwidth]{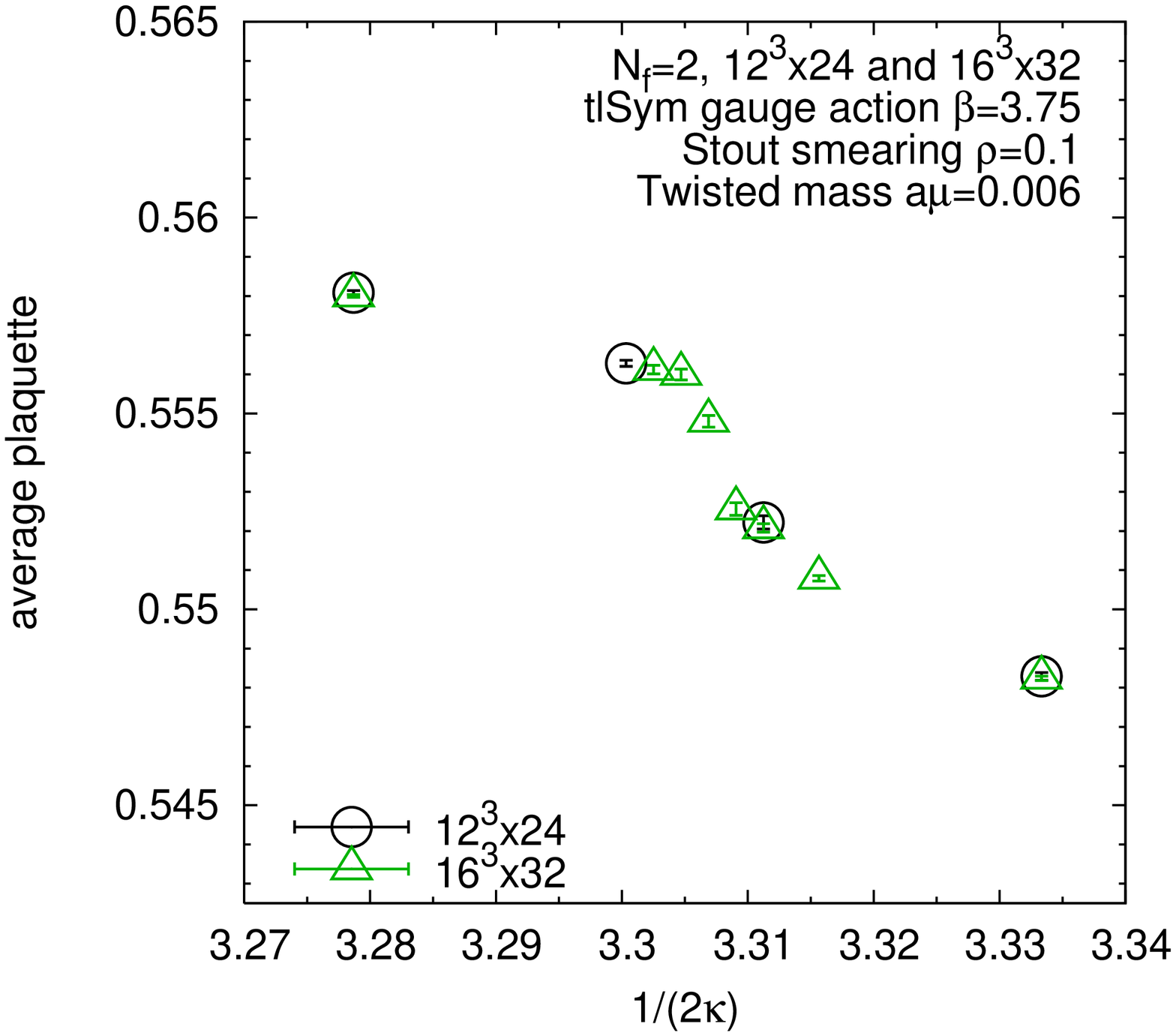}%
\includegraphics[angle=-90, width=.52\textwidth]{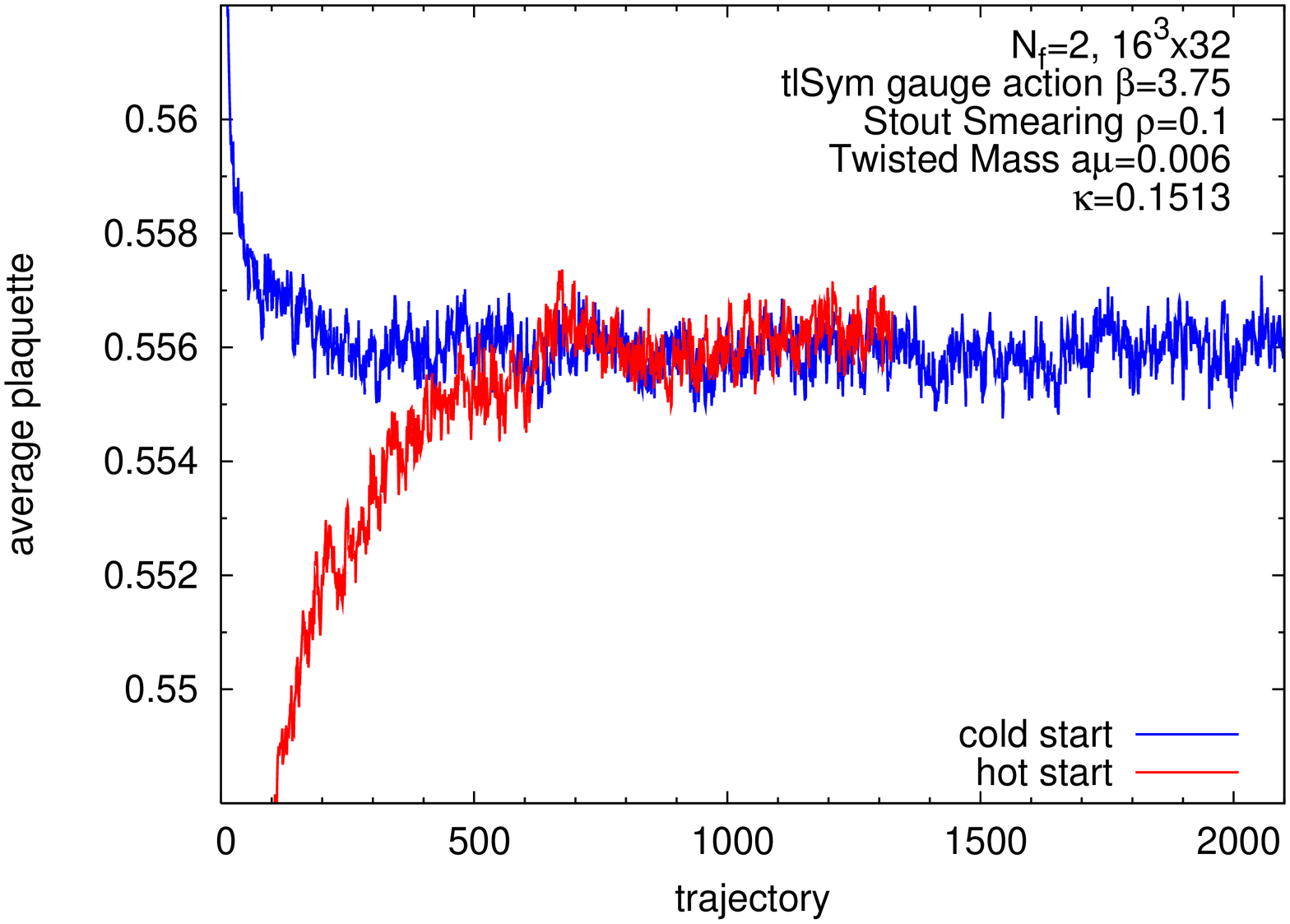}
\end{center}
\vspace*{\closercaption}
\caption{\label{fig:nf2_plaq_Stout}Simulations with Stout smearing. \textit{Left panel:} Average plaquette value on two different lattice sizes. \textit{Right panel:} Monte Carlo history of the plaquette value for hot \textit{(red)} and cold \textit{(blue)} starts.}
\end{figure}

\begin{figure}
\begin{center}
\includegraphics[angle=-90, width=.45\textwidth]{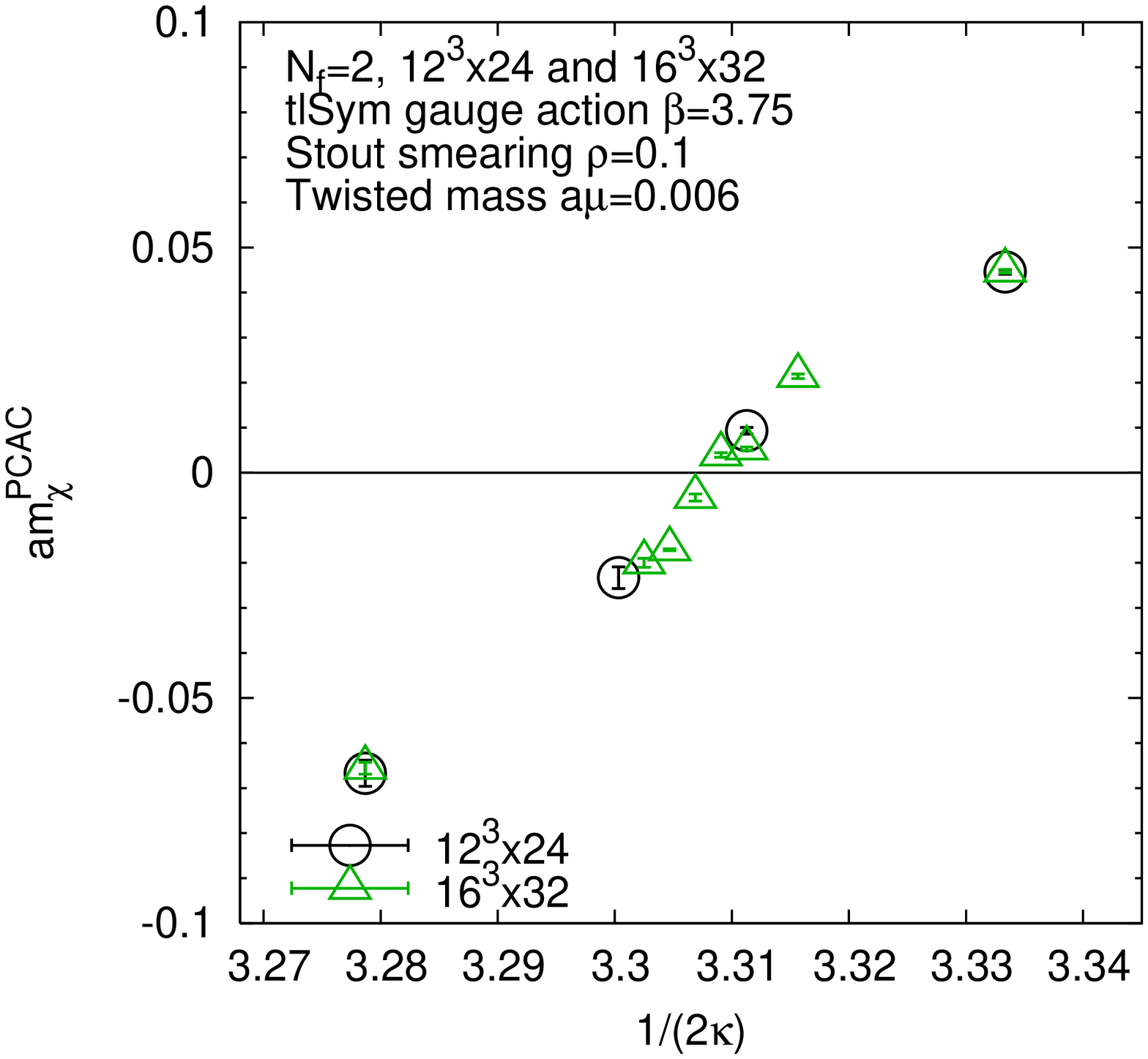}%
\includegraphics[angle=-90, width=.45\textwidth]{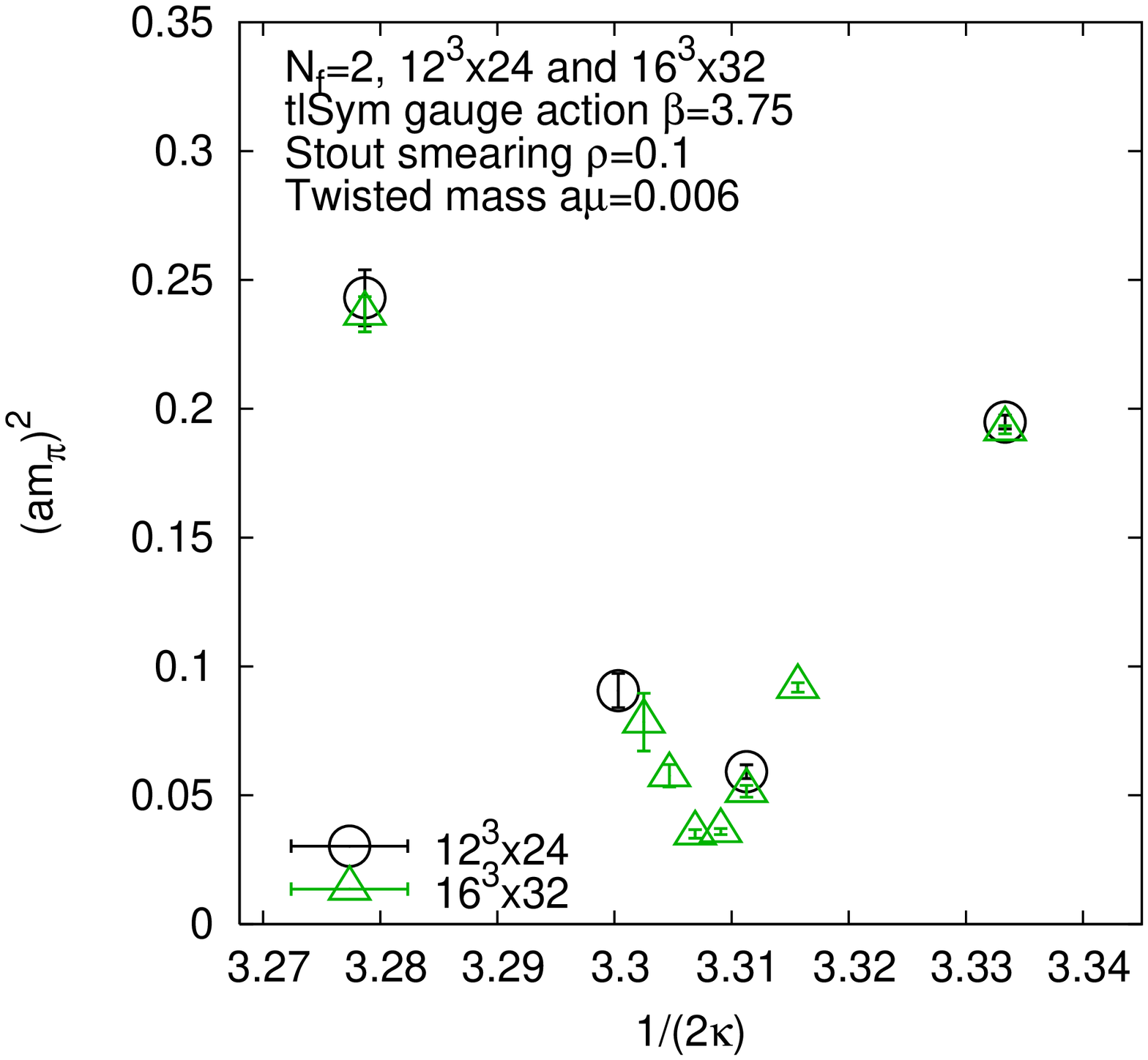}%
\end{center}
\vspace*{\closercaption}
\caption{\label{fig:pcac_pionMass}\textit{Left panel:} PCAC quark mass, \textit{right panel:} squared pion mass using Stout smearing.}
\end{figure}

 Figures~\ref{fig:nf2_plaq_noStout} and \ref{fig:nf2_plaq_Stout} show the average value of the (thin) plaquette without
 and with smearing, respectively. In the case without smearing a jump in the average plaquette value is clearly visible. Here we also observed metastabilities, which show up as differences between runs starting from a random (hot) or ordered (cold) configuration (red circles and blue triangles, respectively, in Fig.~\ref{fig:nf2_plaq_noStout}). In the case of Stout smearing it is unclear whether there is still a phase transition at all, since the left panel of Fig.~\ref{fig:nf2_plaq_Stout} shows a rather smooth dependence of the average plaquette value on the inverse hopping parameter. To examine if metastabilities may still arise with Stout smearing, we started runs from either a random (hot) or ordered (cold) configuration at the same parameters where the hopping parameter was chosen to lie in the region of fastest increase of the average plaquette: $\kappa=0.1513$ or $1/(2\kappa)\approx3.305$; the Monte Carlo histories of the two runs are displayed in the right panel of Fig.~\ref{fig:nf2_plaq_Stout}. One can see that after roughly 500 trajectories both runs thermalized at the same average value for the plaquette giving no evidence to the existence of metastabilities.
 Figure~\ref{fig:pcac_pionMass} shows the (untwisted) PCAC quark
 mass (left panel) and the squared pion mass (right panel) as a function of $\mu_\kappa$.
 The former also shows no clear evidence for the presence of a phase transition, since both branches
 (positive and negative PCAC quark mass) extrapolate to roughly the same
 critical value of $\mu_\kappa$.
 From the latter one can read off that on the volume $L/r_0 \simeq 4$
 a minimal pion mass of $m_\pi r_0 \simeq 0.7$ is easily achieved for
 $a\approx r_0/4$.
 (We use here for setting the scale the Sommer parameter $r_0$.)

%
%

%

On the $24^3\times 48$ lattice in a previous simulation at $\beta=3.8$ \cite{Urbach:Lat07TMreview} without smearing we observed a problematic behaviour in time histories implying very long autocorrelations.   
We are presently repeating this run with one level of Stout smearing ($\rho=0.125$) to see the effect of smearing for such a situation. Although our first
results indicate that smearing helps, it is too early to give a definite conclusion at this point. 
It will also be interesting to
compare physical observables, e.g., $f_\mathrm{PS}$ and $m_\mathrm{PS}$, between the Stout smeared and unsmeared simulations.

\subsection{\label{subsec:nf211}$N_f=2+1+1$}

\begin{figure}
\begin{center}
\hfill
\includegraphics[angle=-90, width=.4\textwidth]{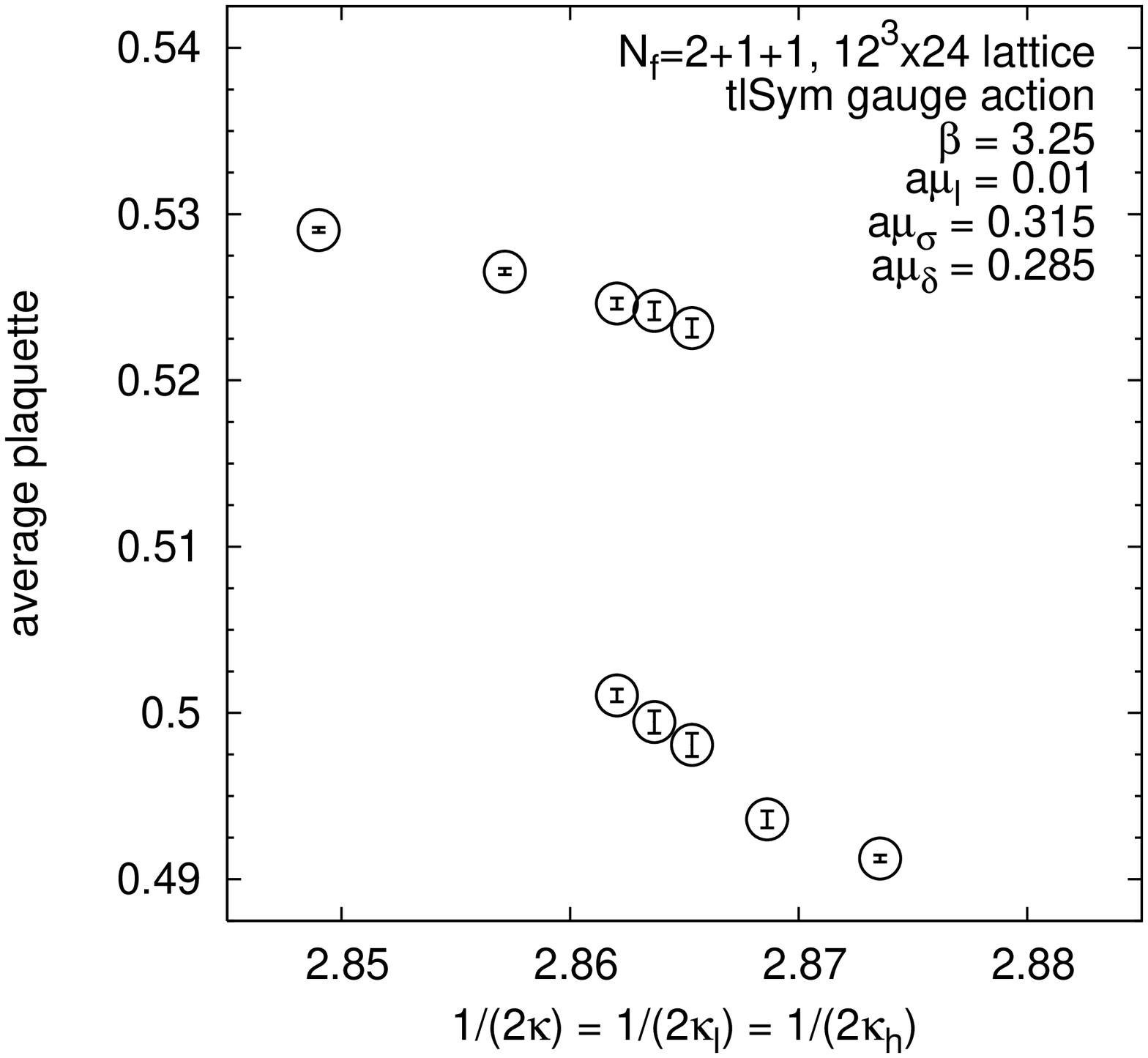}
\hfill
\includegraphics[angle=-90, width=.4\textwidth]{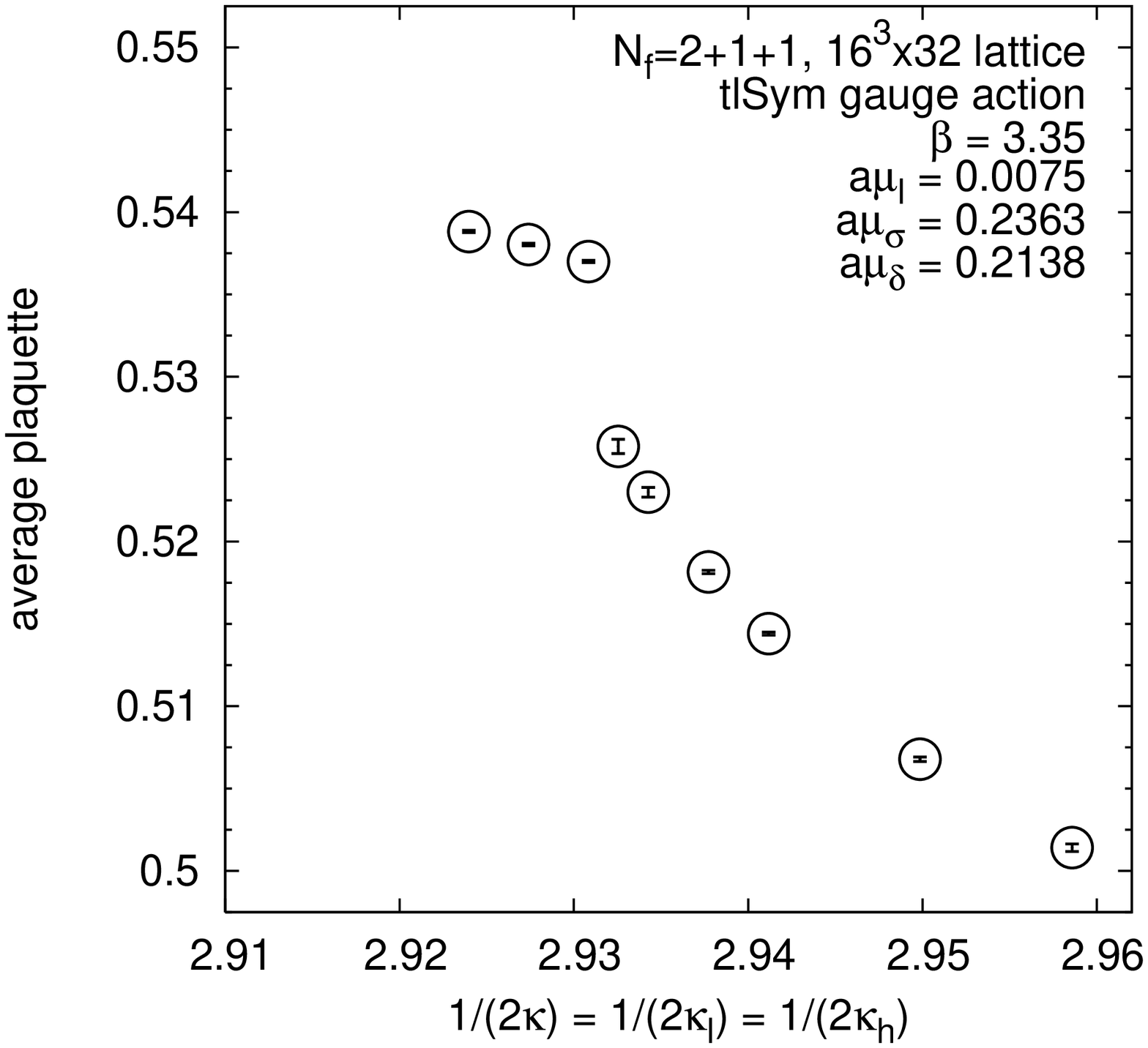}
\hfill$\phantom{.}$
\end{center}
\vspace*{\closercaption}
\caption{\label{fig:nf211}Average plaquette with $N_f=2+1+1$ flavours on $12^3\times24$, $\beta=3.25$ \textit{(left panel)} and $16^3\times32$, $\beta=3.35$ \textit{(right panel)} lattices, see \cite{Chiarappa:2006ae}.}
\end{figure}

 Recently, the possibility of adding the strange quark in dynamical twisted mass
 simulations has been tried following the lines of \cite{Frezzotti:2003xj} by
 introducing a mass splitting term in the heavier doublet, see
 Eq.~(\ref{eq:tm_heavyDoublet}).
 In that way not only a strange quark will be added but also the much heavier
 charm quark is taken into account.
 For first numerical results see \cite{Chiarappa:2006ae}, where an important
 conclusion is that the extra dynamical quarks strengthen the first order phase
 transition.
 As an example, in Fig.~\ref{fig:nf211} we show the jump in the average plaquette
 for two different lattice spacings at a fixed physical volume.
 On the coarser lattice spacing (left panel) again metastabilities show up.
 At the finer lattice spacing (right panel) there are no more metastabilities
 but there is still a considerable ``jump'' in the average plaquette.
The findings from our Stout smeared run for $N_f=2$ suggest that smearing could substantially help in the case of $N_f=2+1+1$.
%


\section*{Conclusions \& Outlook}

 The conclusion of testing Stout smearing with twisted mass Wilson quarks is that
 the first order phase transition at non-zero lattice spacing becomes weaker as
 a result of smearing.
 Therefore moderate Stout smearing can be an option---in particular for future
 numerical simulations in the twisted mass formalism with dynamical $u$-, $d$-, $s$- and
 $c$-quarks.

\noindent\textbf{Acknowledgements.} We would like to thank S.~D\"urr for discussions. The numerical simulations were performed on the QCDOC at Edinburgh, the JUMP at Forschungszentrum J\"ulich, and the Scotgrid. This work has been supported in part by the EU Integrated Infrastructure Initiative Hadron Physics (I3HP) under contract RII3-CT-2004-506078\@. E.S.\ was supported by the U.S.\ Dept.\ of Energy under contract DE-AC02-98CH10886.

%

\bibliography{references}

\bibliographystyle{JHEP-2} 

\end{document}